\documentstyle[twoside,fleqn,espcrc2,psfig]{article}
\newcommand{\AmS}{{\protect\the\textfont2
  A\kern-.1667em\lower.5ex\hbox{M}\kern-.125emS}}

\title{Suppressing Curvature Fluctuations in Dynamical
 Triangulations}

\author{M.J.~Bowick$^{\rm a}$, S.M.~Catterall$^{\rm a}$ 
  {\rm and {\bf G.~Thorleifsson}}\address{Department of Physics, 
  Syracuse University, Syracuse, NY 13244-1130, USA}} 
        
\begin{document}

\begin{abstract}
We study numerically the dynamical triangulation formulation
of two-dimensional quantum gravity using a restricted 
class of triangulation, so-called {\it minimal triangulations},
in which only vertices of coordination number 
5, 6, and 7 are allowed \cite{r1}. A real-space
RG analysis shows that for pure gravity (central charge $c=0$)
this restriction does not affect the critical behavior of the
model.  Furthermore, we show that the critical behavior of
an Ising model coupled to minimal dynamical triangulations ($c=1/2$)
is still governed by the KPZ-exponents. 
\end{abstract}

\maketitle

\section{INTRODUCTION}

In the discrete formulation of two-dimensional quantum 
gravity, known as {\it dynamical triangulations}, the          
micro{--}canonical, or fixed area, partition function
is defined as the sum over all possible ways of gluing
$A$ equilateral triangles together to form a 
piecewise linear manifold (with an appropriate topology):
\begin{equation}
\label{*11}
Z_{\cal T}(A) \;=\; \sum_{T\in{\cal T}} Z_M(T) .
\end{equation}
In the case of matter coupled to gravity, the triangulations $T$
are weighted with the partition function $Z_M$ for the corresponding
matter fields.

In Eq.~(\ref{*11}) ${\cal T}$ represents the class of 
triangulations (of area $A$) included in the partition function.
Different classes correspond to different discretization 
of the manifolds.  Two commonly used classes are:
{\it combinatorial} triangulations ${\cal T}_C$, and
{\it degenerate} triangulations ${\cal T}_D$.  In the
former a vertex is not allowed to connected to itself, nor can
any two vertices be connected by more than one link {---}
this excludes tadpole and self-energy 
diagrams in the dual graph.  This restriction is
eased for degenerate triangulations.  Clearly
${\cal T}_C \subset {\cal T}_D$.  In cases where the
model Eq.~(\ref{*11}) has been solved, 
it has been shown that $Z_{{\cal T}_C}$
and $Z_{{\cal T}_D}$ are in the same universality class
\cite{r2}.

But how general is this universality, i.e.\ how much restriction
can be imposed on the triangulations before the critical behavior
of the model changes 
{---} clearly including only one triangulation would produce
different result.  In the work presented here, we investigate
this by considering a new class of triangulations ${\cal T}_M$,
the class of {\it minimal triangulations},
in which only vertices with coordination number 
5, 6, and 7 are allowed.  
This is as far as we can go in suppressing curvature
fluctuations, while still retaining the fluid nature of 
the surfaces.  We will demonstrate, by studying the critical
behavior of pure gravity and an Ising matter coupled to gravity,
that this restriction does not change the critical
behavior of the model.

An additional motivation for studying this class of 
triangulations is that they more closely resemble conventional condensed
matter systems.  Previously studied models of dynamical 
triangulations allow vertices with arbitrary high
curvature, whereas a real system places a cut-off 
on the number of interaction for a single particle. 
That minimal dynamical triangulations
are in the usual universality class 
of $2D$ gravity, implies that the KPZ-exponents
might be realized in a real system (yet to be discovered).

\section{PURE GRAVITY}

We start by investigating this model in the absence of
matter ($Z_M = 1$).  To determine the critical
behavior we study the fractal structure
of the surfaces.  More precisely we measure the
string susceptibility exponent $\gamma_s$, which governs
the critical behavior of the grand canonical
partition function $Z(\mu) \sim (\mu_c - \mu)^{2-\gamma_s}$.
In numerical simulations, $\gamma_s$ is obtained 
from the size distribution of
minimal neck baby universes $n(B)$ \cite{r3};
\begin{equation}
\label{*22}
n_A(B) = \left [ B(A-B) \right ]^{\gamma_s-2}.
\end{equation}
For pure gravity $\gamma_s = -\frac{1}{2}$.

\begin{figure}[t]
\psfig{figure=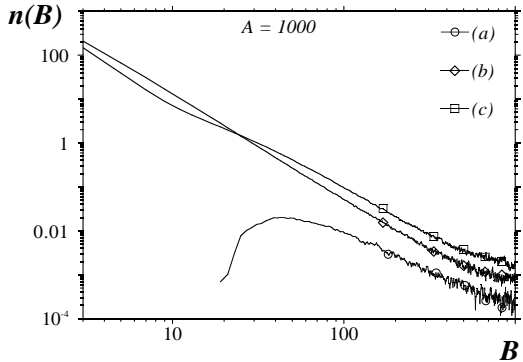,width=2.8in}
\caption{The measured distributions of baby universes for
 {\it (a)} the minimal class of triangulations, {\it (b)}
 combinatorial triangulations and {\it (c)} triangulations
 obtained by node decimation from {\it (a)}.  All measurements
 are for triangulations of 1000 vertices.}
\label{fig1}
\end{figure}

We have measured the distribution $n(B)$ for 
minimal triangulations,  this is shown in Fig.~\ref{fig1}
(curve $a$).  For comparison we also show the
corresponding distribution measured on combinatorial
triangulations (curve $b$). In addition Fig.~\ref{fig1}
includes a distribution 
measured on triangulations obtained from minimal triangulations
via node decimation.  We will discuss this later.

Although, for large baby universes, the slope of the distributions
$a$ and $b$ is similar, Fig.~\ref{fig1} clearly
shows much bigger finite size effects for minimal
triangulations.  This is easily understood, as the
restrictions on minimal triangulations effectively smoothens the
surfaces locally --- in fact there is a lower cut-off on the
size of the baby universes.  An estimate of $\gamma_s$
for the minimal triangulations yields 
$-0.64(5)$ for $A=1000$ and $-0.53(3)$ for $A=4000$, compared
to $\gamma_s = -0.501(4)$ for combinatorial triangulations
and the exact value $-\frac{1}{2}$.

The same result is also obtained by 
using a recently proposed real-space RG method for dynamical
triangulations; {\it node decimation} \cite{r4}.  
Applying node decimation we have blocked the minimal
triangulations repeatedly, down to 250 nodes, 
using a blocking factor of 2.
Note that under blocking, the minimal triangulations flow
into the wider class of combinatorial triangulations.
We then measure the distribution $n(B)$ on the ensemble
of blocked triangulations.  This is shown in Fig.~\ref{fig1}.
Clearly the distribution obtained from minimal triangulations
via blocking is much closer to the distribution for combinatorial
triangulations than the original --- this implies that
the finite size effects are reduced in the blocking.
This also implies that under blocking the model flows towards a
non-trivial fixed point, corresponding to $2D$ gravity.

The reductions of finite size effects is also apparent in
the value of $\gamma_s$, measured on the blocked triangulations. 
This is shown in Table~\ref{tab1}.  Under blocking the
measured value of $\gamma_s$ approaches the exact value.

\begin{table}
\begin{center}
\caption{Measured values of $\gamma_s$ for minimal
dynamical triangulations (without matter) after
applying varying levels of node decimation with
a blocking factor of $b = 2$.}
\begin{tabular}{ccccc} \hline
 & \multicolumn{2}{c}{$A^{(0)} = 1000$}
 & \multicolumn{2}{c}{$A^{(0)} = 4000$}  \\
$A^{(k)}$ &  $k$  &  $\gamma_s$  &  $k$  &  $\gamma_s$   \\ \hline
4000    &       &              &  0    &  -0.530(26)   \\
2000    &       &              &  1    &  -0.544(32)   \\
1000    &  0    &  -0.644(48)  &  2    &  -0.530(18)   \\
500     &  1    &  -0.619(26)  &  3    &  -0.504(9)    \\
250     &  2    &  -0.574(49)  &  4    &  -0.478(36)   \\ \hline \hline
\end{tabular}
\label{tab1}
\end{center}
\end{table}

\begin{table*}[hbt]
\setlength{\tabcolsep}{1.5pc}
\newlength{\digitwidth} \settowidth{\digitwidth}{\rm 0}
\catcode`?=\active \def?{\kern\digitwidth}
\begin{center}
\caption{The measured critical exponents. $\nu d_H$ and $\alpha$
 are obtained from the scaling of the peaks in $\partial g_r/
 \partial \beta $ and $C_V$ respectively.  $\beta$ and $\gamma$
 are measured from scaling at $\beta_c$, in which case the errors
 are dominated by the uncertainty in the location of $\beta_c$.}
\begin{tabular}{ccccc} \hline
$A$   & $\nu d_H$ & $\alpha$ & $\beta$ & $\gamma$  \\ \hline
250-8000  & 2.857(24) & -0.806(41) & 0.474(20) & 2.117(54) \\
500-8000  & 2.890(33) & -0.922(58) & 0.488(21) & 2.095(52) \\
1000-8000 & 2.907(25) & -0.977(87) & 0.500(18)
 & 2.070(55) \\ \hline
Onsager   & 2         & 0(log)     & 1/8        & 7/4 \\
KPZ       & 3         & -1         & 1/2        & 2   \\
\hline \hline
\end{tabular}
\label{tab2}
\end{center}
\end{table*}

\section{THE ISING MODEL}

We have also investigated the critical behavior of an Ising
model coupled to minimal triangulations, for 
lattice sizes $A = 250$ to 8000.  In this case
\begin{equation}
\label{*31}
Z_M(\beta) = \sum_{\{\sigma_i\}} {\rm e}^{\beta \sum_{<ij>}
\sigma_i \sigma_j },
\end{equation}
where $\sigma_i$ is an Ising spin placed on vertex $i$.
The Ising spins were updated using a Swendsen-Wang
cluster algorithm and approximately $10^7$ sweeps
performed per data-point.

We must first determine the infinite volume critical coupling
$\beta_c$.  This is done by
locating the peaks in the specific heat $C_V$ and the derivate
of Binders cumulant $\partial g_r / \partial \beta$;
both are expected to approach $\beta_c$ as
\begin{equation}
\label{*32}
| \beta_c - \beta_c(A) | \sim A^{-1/\nu d_H}.
\end{equation}
The fit to Eq.~\ref{*32} is made easier by an independent
determination of $\nu d_H$ --- the height of
$\partial g_r / \partial \beta$
scales like $A^{1/\nu d_H}$.  This procedure
yields $\beta_c = 0.2663(3)$ for the Ising model coupled
to minimal dynamical triangulations.

Using this estimate of $\beta_c$, we then determine
other critical exponents using finite size scaling.
We have the magnetization; $M \sim A^{-\beta/\nu d_H}$,
the magnetic susceptibility; $\chi \sim A^{\gamma /\nu d_H}$,
and the specific heat; $C_V \approx c_0 + c_1 A^{\alpha/\nu d_H}$.
The exponents are shown in Table~2 together with the
KPZ and Onsager predictions.  We indicate
the finite size effects by imposing different lower cut-off's on the
lattice size used.
But the conclusion is clear --- the measured exponents agree very
well with the KPZ-exponents.

\section{DISCUSSION}

In this paper, we have investigated the critical behavior of a 
so-called minimal dynamical triangulation model of $2D$ gravity.
In this model only vertices of coordination numbers 5, 6, and 7 are
allowed --- this effectively suppresses the high curvature fluctuations.
By studying the fractal structure of pure gravity,
and the critical properties of an Ising model coupled to
gravity, we conclude that the critical behavior is not affected
by this restriction to minimal triangulations --- the model
is in the same universality class as
with combinatorial or degenerate triangulations.
This implies that large curvature fluctuations do not play an
important role in determining the continuum structure of
$2D$ quantum gravity.  

This result agrees with 
recent studies of a matrix-model formulation of $R^2$-gravity,
where similarly the $R^2$-term can be used suppress
the curvature fluctuations locally.  In \cite{r5} it was
shown that the $R^2$ operator is irrelevant in the 
continuum limit --- at large length scales the model always
reduces to that of pure gravity.

\end{document}